\documentclass[prl,aps,superscriptaddress,twocolumn]{revtex4}
\usepackage{rotating}
\usepackage{graphicx}% Include figure files
\usepackage{dcolumn}% Align table columns on decimal point
\usepackage{bm}% bold math
\usepackage{amssymb}
\begin{document}

\title{Critical Properties of the Kitaev-Heisenberg Model}

\author{Craig C. Price}
\author{Natalia B. Perkins}

\affiliation{
Department of Physics,
University of Wisconsin,
1150 University Ave.,
Madison, Wisconsin 53706, USA
}

\begin{abstract}
We  study  the critical properties  of the Kitaev-Heisenberg (KH)  model on the honeycomb lattice at finite temperatures that might describe the  physics of the quasi two-dimensional (2D) compounds, Na$_2$IrO$_3$  and  Li$_2$IrO$_3$.
The model undergoes two phase transitions as a function of temperature.
At low temperature, thermal fluctuations induce magnetic long-range order by the order-by-disorder mechanism.
This magnetically ordered state with a spontaneously broken $Z_6$ symmetry persists up to a certain critical temperature.
We find that there is an intermediate phase between the low-temperature, ordered phase  and the high-temperature, disordered phase.
Finite-sized  scaling analysis suggests that the intermediate phase is a critical Kosterlitz-Thouless (KT)  phase with continuously variable exponents.
We argue that the intermediate phase has been observed  above the low-temperature, magnetically ordered phase in Na$_2$IrO$_3$, and also likely exists in  Li$_2$IrO$_3$.
\end{abstract}

\maketitle
%\section{Introduction}
{\it Introduction.}
The Ir-based  transition metal oxides, in which the orbital degeneracy is accompanied by a strong relativistic spin-orbit coupling (SOC), have recently attracted a lot  of theoretical and experimental attention \cite{nakatsuji06,okamoto07,kim09,gegenwart10,gegenwart12,liu11,choi12,takagi11}.
This is because the strong SOC creates a different, and frequently novel,  set of magnetic and orbital states due to the unusual anisotropic exchange interactions between localized moments which are in turn determined by the combination of spin and lattice symmetries.
%The interest in these systems is due to a variety of  magnetic and orbital states different from that  in the systems with weak SOC.
%The novel and peculiar features of these systems are due to unusual  anisotropic exchange interactions between  localized moments that are determined by a combination of spin and lattice symmetries.
The spin-orbital models that describe the low-energy physics of iridium systems often include anisotropic terms that do not reduce to the conventional easy-plane and easy-axis anisotropies because they involve  the products of different components of  multiple spin operators.
These terms are responsible  for exotic  Mott-insulating states~\cite{kim09}, topological insulators \cite{Shitade09,Pesin10},  spin-orbital liquid states \cite{nakatsuji06,okamoto07}, and non-trivial long-range magnetic orders \cite{kim09,gegenwart10,liu11}.

A prominent example of  such an anisotropic spin-orbital model is  the KH  model on the honeycomb lattice \cite{jackeli09,jackeli10} which likely describes the low-energy physics of the quasi 2D  compounds, Na$_2$IrO$_3$  and  Li$_2$IrO$_3$.
In these compounds, Ir$^{4+}$ ions  are in a  low spin $5d^5$ configuration   and form  weakly coupled hexagonal layers \cite{gegenwart10,liu11,takagi11}.
Due to strong SOC, the atomic ground state is a doublet where the spin and orbital angular momenta of Ir$^{4+}$ ions  are coupled into  $J_{\rm eff} = 1/2$.
It was suggested \cite{jackeli09,jackeli10} that the interactions between these effective moments can be described by a  spin Hamiltonian containing two competing nearest neighbor  (NN) interactions:
%In refs.  it has been suggested  that the interactions between these effective spin one-half moments  can be described by a  spin Hamiltonian containing two competing nearest neighbor  (NN) interactions:
an isotropic antiferromagnetic  (AF) Heisenberg exchange interaction and a highly anisotropic ferromagnetic (FM)   Kitaev exchange interaction \cite{kitaev06}.
This competition can be described with the parameter, $0\leq\alpha\leq 1$, which sets the relative strength of these two interactions.
At $\alpha =0$,  the coupling
%between the moments of Ir ions
corresponds to the AF Heisenberg interaction, and at $\alpha =1$, it corresponds to the Kitaev interaction.

This model immediately attracted a lot of attention; several theoretical studies were published in the last few years~\cite{jackeli10,jiang11,reuther11,trousselet11} on both the ground state and its properties at finite temperature.
The ground state phase diagram of the KH model exhibits three distinct phases:
the   AF N\'{e}el  phase  for small $\alpha\in(0.,0.4)$,
the stripy AF phase for intermediate $\alpha \in (0.4,0.86)$,
and the disordered spin-liquid phase at large $\alpha\in (0.86,1.)$.
While the phase transition between the N\'{e}el and the stripy phase appears to be discontinuous, numerical studies including density matrix renormalization group~\cite{jiang11} and exact diagonalization results~\cite{jackeli10} suggest that the transition between the spin liquid and the stripy state is continuous or weakly first-order.
%Another important feature of the ground state properties of the model is that in all magnetically ordered phases of the phase diagram the order parameter points along one of the cubic axes which is selected  by quantum  fluctuations.
Additionally, quantum fluctuations select all of the magnetically ordered phases to have the order parameter point along one of the cubic axes.

%
%Perhaps, per Referee A's advice, we should move more of the next three paragraphs to the results section. This might also help fit the space limit.
%

In this Letter, we discuss finite temperature properties of the KH model on the honeycomb lattice.
A first step in this direction was made in Ref.~\cite{reuther11}, where the critical ordering scale for the magnetically ordered states  was analyzed using a pseudofermion functional renormalization group  approach.
Here we present  numerical results obtained using Monte Carlo (MC) simulations.
We study the classical  KH model because the corresponding quantum model has a sign problem precluding quantum MC analysis and also because the existence of long-range order  at low temperatures in Na$_2$IrO$_3$ and  in Li$_2$IrO$_3$ indicates that quantum fluctuations are not dominating in these materials~\cite{gegenwart10,gegenwart12,liu11,choi12,takagi11}.

We show that the thermal fluctuations  of classical spins give rise to two distinct temperature dependent effects.
At low temperature they  predominantly act as the source of the order-by-disorder phenomenon and select collinear magnetic order where the spins are oriented along one of the cubic directions.
There are six possible  ordered states, one of which is spontaneously chosen by the system.
At high temperatures, when $T$ is larger than any energy scale in the system, the fluctuations destroy any order putting the KH model into a three dimensional paramagnetic state.
The main goal of our study is to see how these two phases are connected.

We argue that the classical KH model effectively behaves like a six-state clock model~\cite{jose77,isakov03,chern12,ortiz12} % and has remarkable properties at finite temperatures.
and that it undergoes two continuous phase transitions as a function of temperature separating three phases: a low-T ordered phase, an intermediate  critical phase, and a high-T disordered phase.
The critical phase  has an emergent, continuous $U(1)$ symmetry which is fully analogous to  the low-T phase of the XY model, a well-known KT phase of critical points with floating exponents and algebraic correlations.
Here we  present numerical data only for $\alpha=0.25$ and $\alpha=0.75$ since these values likely characterize the ratio between  the AF Heisenberg interaction and the Kitaev interaction in Na$_2$IrO$_3$  and  Li$_2$IrO$_3$.
%The full finite-temperature phase diagram  for the KH model will be published elsewhere \cite{unpublished}.
However, we note that recent inelastic neutron scattering measurements on Na$_2$IrO$_3$ have shown that the KH model alone is insufficient to describe the magnetic properties of this compound~\cite{choi12}.
It has been  demonstrated that it is essential to  include substantial further-neighbor exchanges to describe both the zigzag ground state and the excitation spectrum in Na$_2$IrO$_3$.
The  full finite-temperature phase diagram  for the KH model  with second and third neighbor exchange interactions  will be  published elsewhere \cite{unpublished}.

%%%%%%%%%%%%%%%%%%%%%%%%%%%%%%%%%%%%%%%%%%%%%%%%%%%%%%%%%%%%%%%%%%%%%%%%%%%%%%%%
\begin{figure}
\includegraphics[width=0.65\columnwidth]{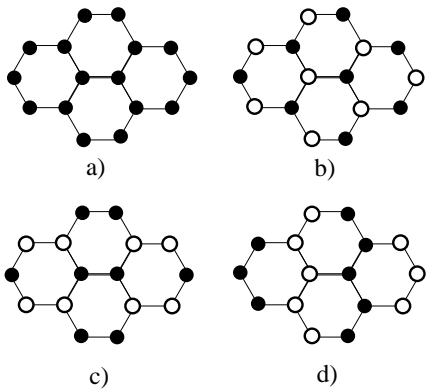}
\caption{
Four possible magnetic configurations:
(a) the FM ordering;
(b) the  two-sublattice, AF N\'{e}el order;
(c) the stripy  order;
(d) the zigzag order.
Open and filled circles correspond to up and down  directions of spins.
}
\label{fig:orders}
\end{figure}
%%%%%%%%%%%%%%%%%%%%%%%%%%%%%%%%%%%%%%%%%%%%%%%%%%%%%%%%%%%%%%%%%%%%%%%%%%%%%%%%

{\it The Model.}
The classical version of the KH model which describes the interactions  among  the $ J=1/2$ degrees of freedom of Ir$^{4+}$ ions reads as
%%%%%%%%%%%%%%%%%%%%%%%%%%%%%%%%%%%%%%%%%%%%%%%%%%%%%%%%%%%%%%%%%%%%%%%%%%%%%%%%
\begin{eqnarray}
\label{ham}
    \mathcal{H} =-J_K \sum_{\langle ij \rangle_\gamma} S_i^{\gamma}S_j^{\gamma}+ J_H\sum_{\langle ij \rangle} {\bf S}_i{\bf S}_j~.
\end{eqnarray}
%%%%%%%%%%%%%%%%%%%%%%%%%%%%%%%%%%%%%%%%%%%%%%%%%%%%%%%%%%%%%%%%%%%%%%%%%%%%%%%%
where the spin quantization axes are taken along the cubic axes of the IrO$_6$ octahedra.
$\gamma=x,y,z$ denotes the three bonds of the honeycomb lattice.
The exchange constants, $J_K=2\alpha$ and $J_H=1-\alpha$, correspond to the  Kitaev and   Heisenberg  interactions which can be derived from a  multiorbital Hubbard Hamiltonian \cite{jackeli10}.
%%%%%%%%%%%%%%%%%%%%%%%%%%%%%%%%%%%%%%%%%%%%%%%%%%%%%%%%%%%%%%%%%%%%%%%%%%%%%%%%
\begin{figure}
\includegraphics[width=0.2726\columnwidth]{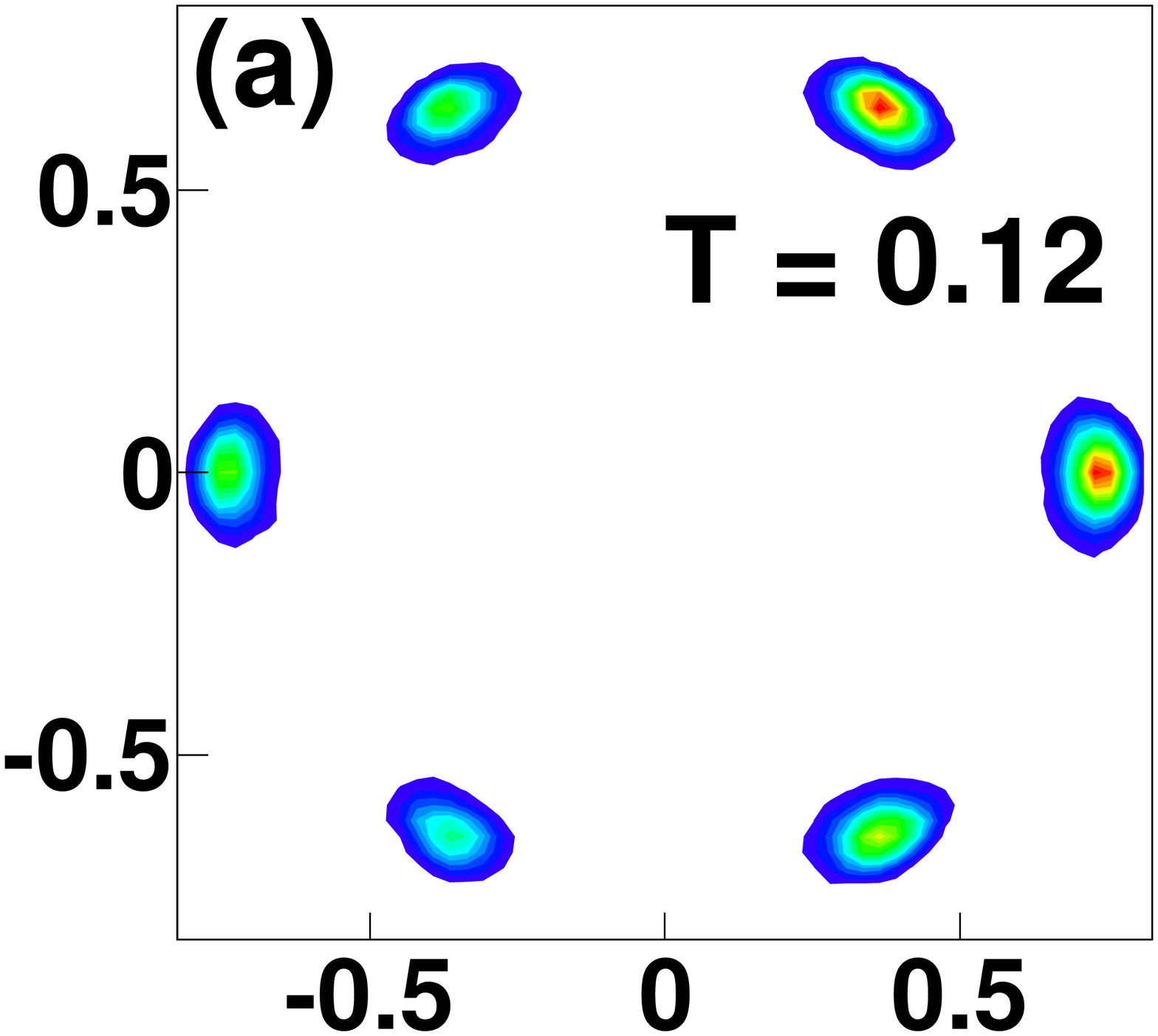}
\includegraphics[width=0.225\columnwidth]{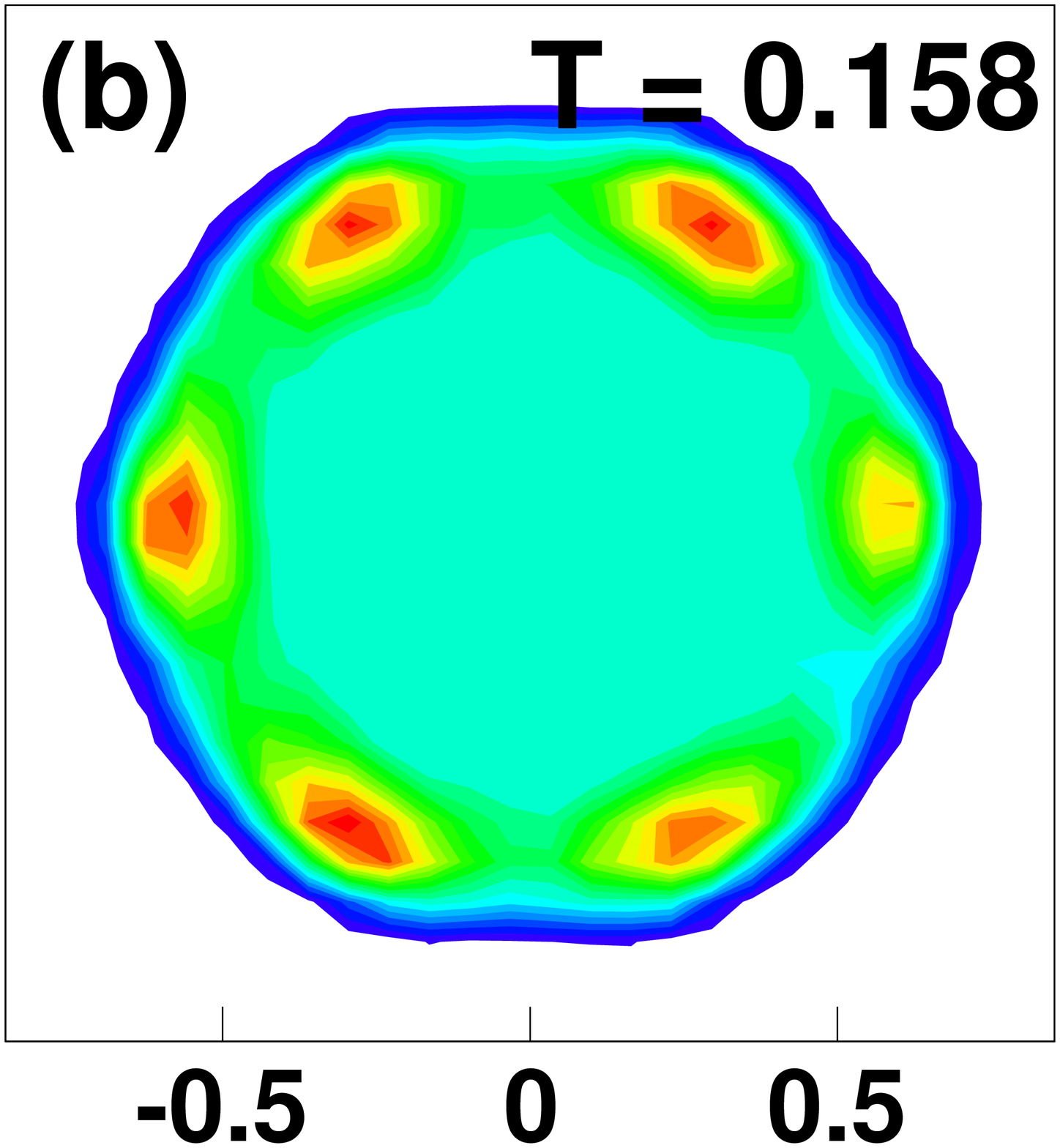}
\includegraphics[width=0.225\columnwidth]{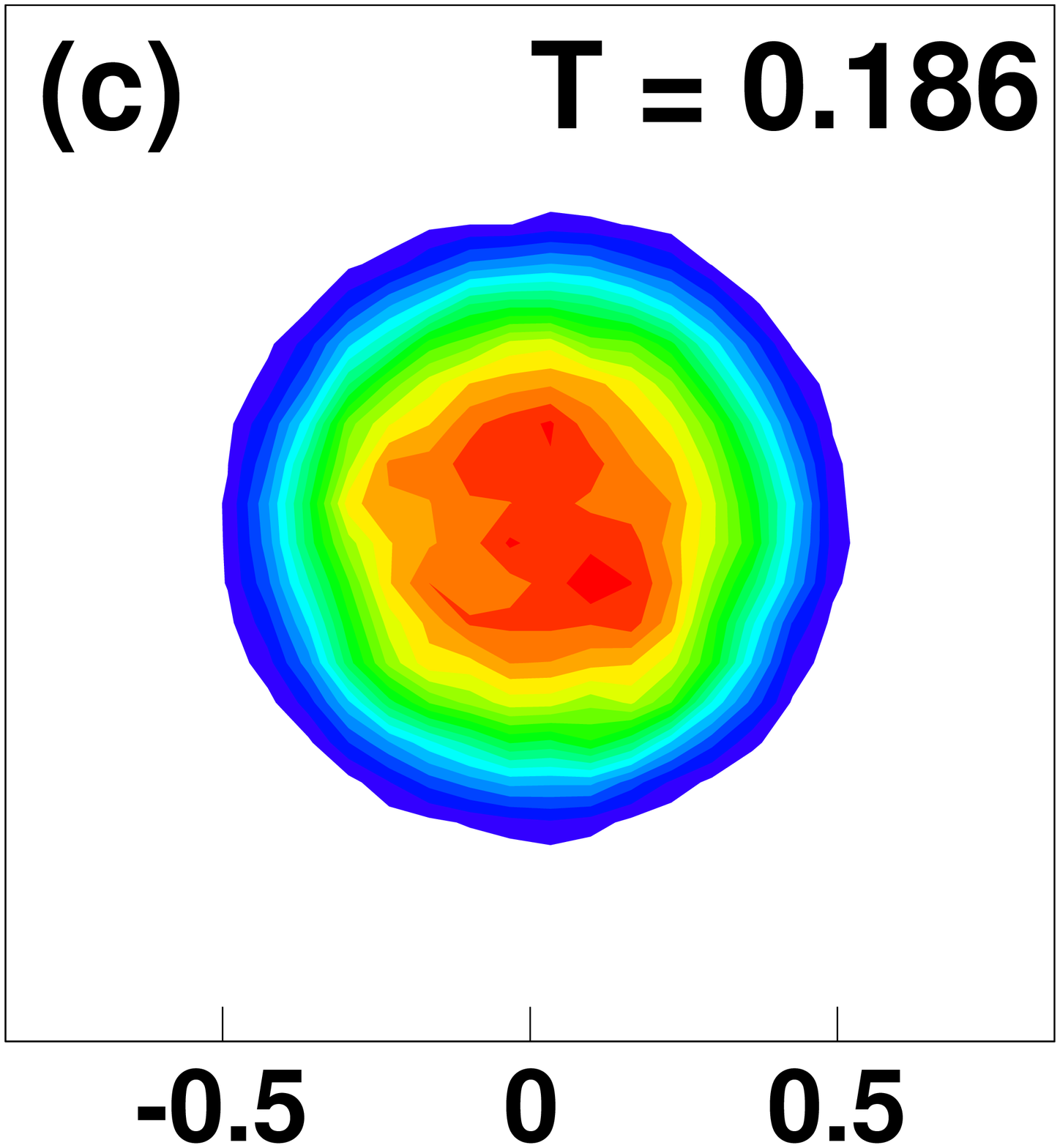}
\includegraphics[width=0.225\columnwidth]{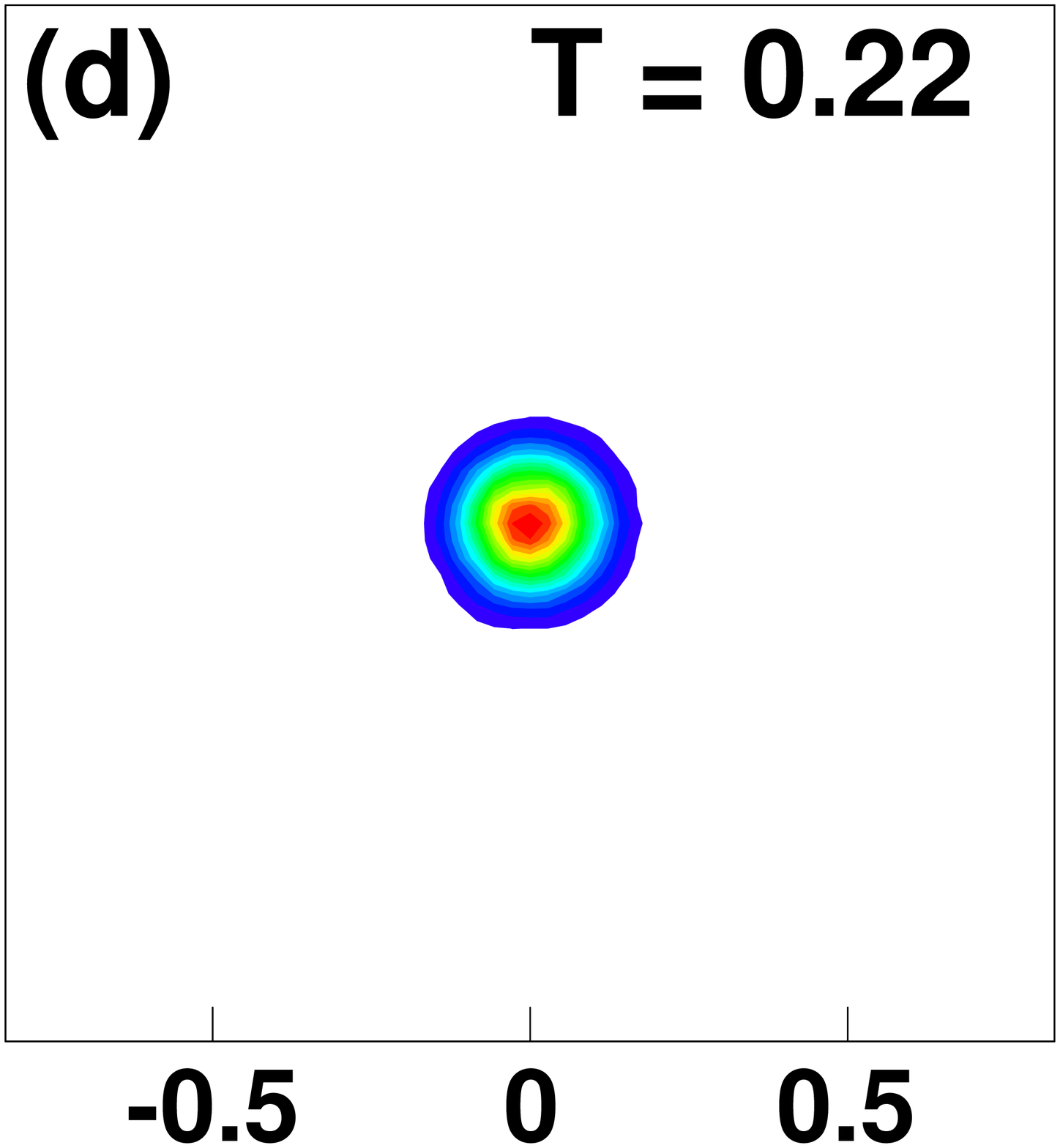}
\includegraphics[width=0.2726\columnwidth]{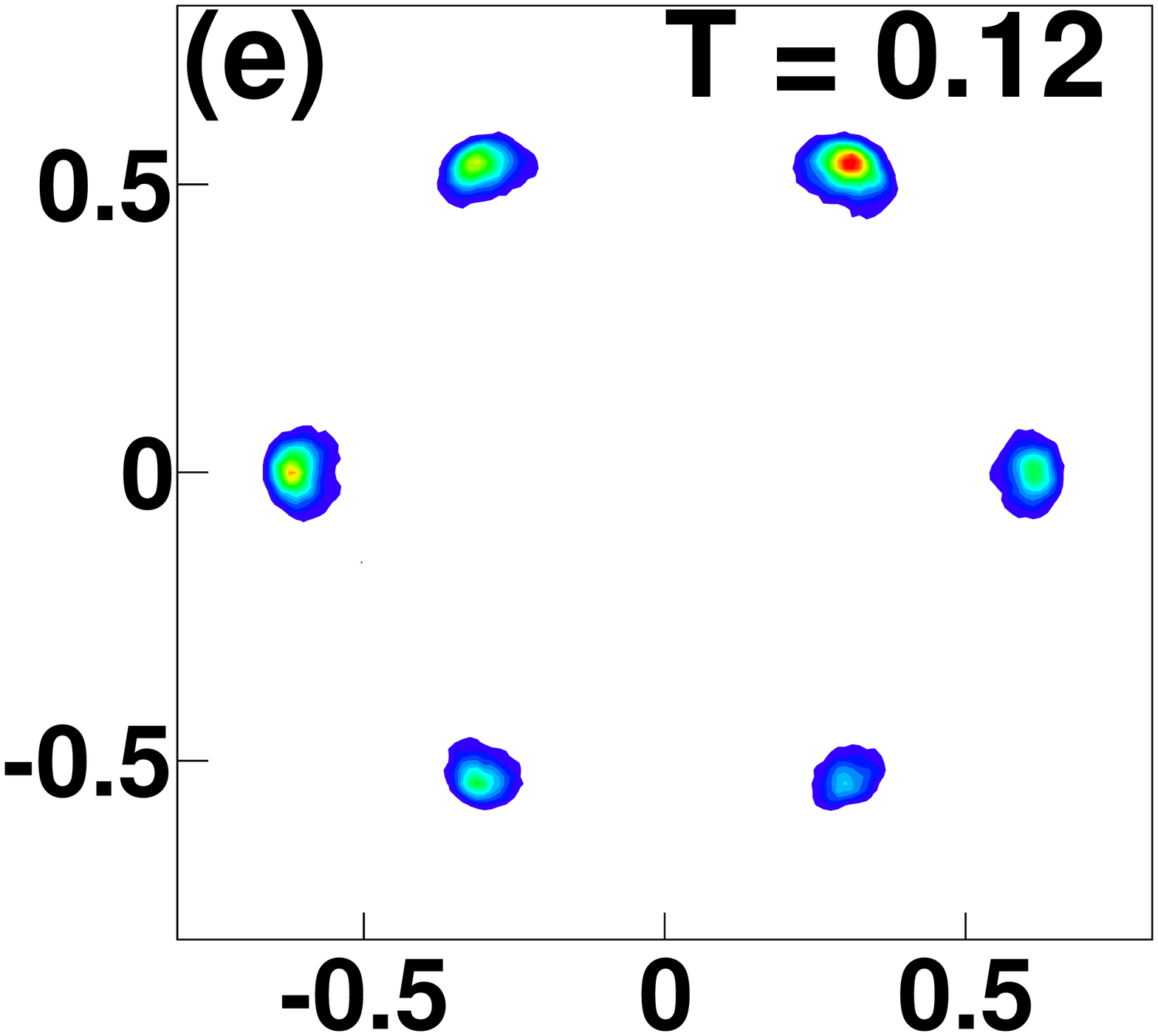}
\includegraphics[width=0.225\columnwidth]{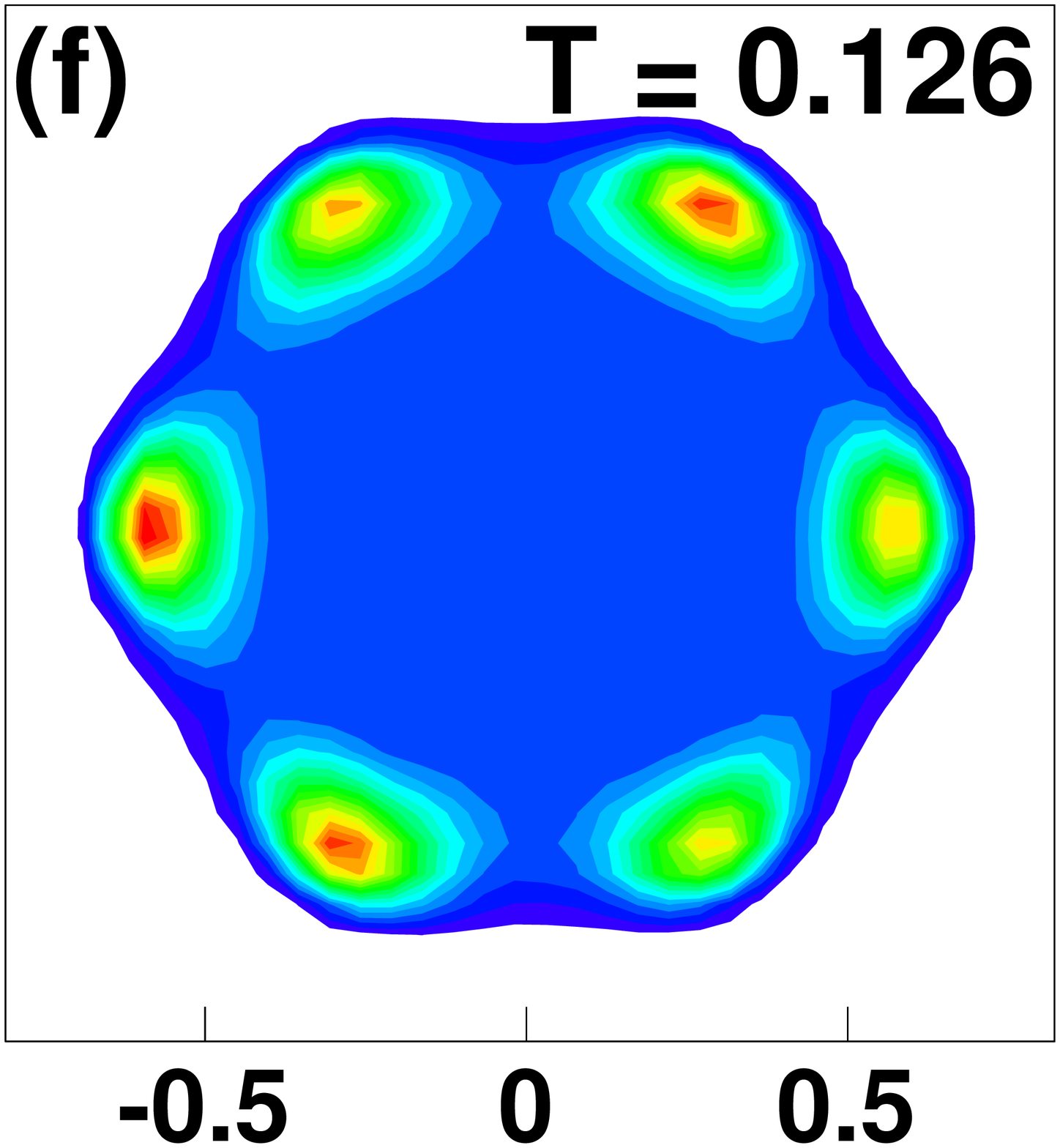}
\includegraphics[width=0.225\columnwidth]{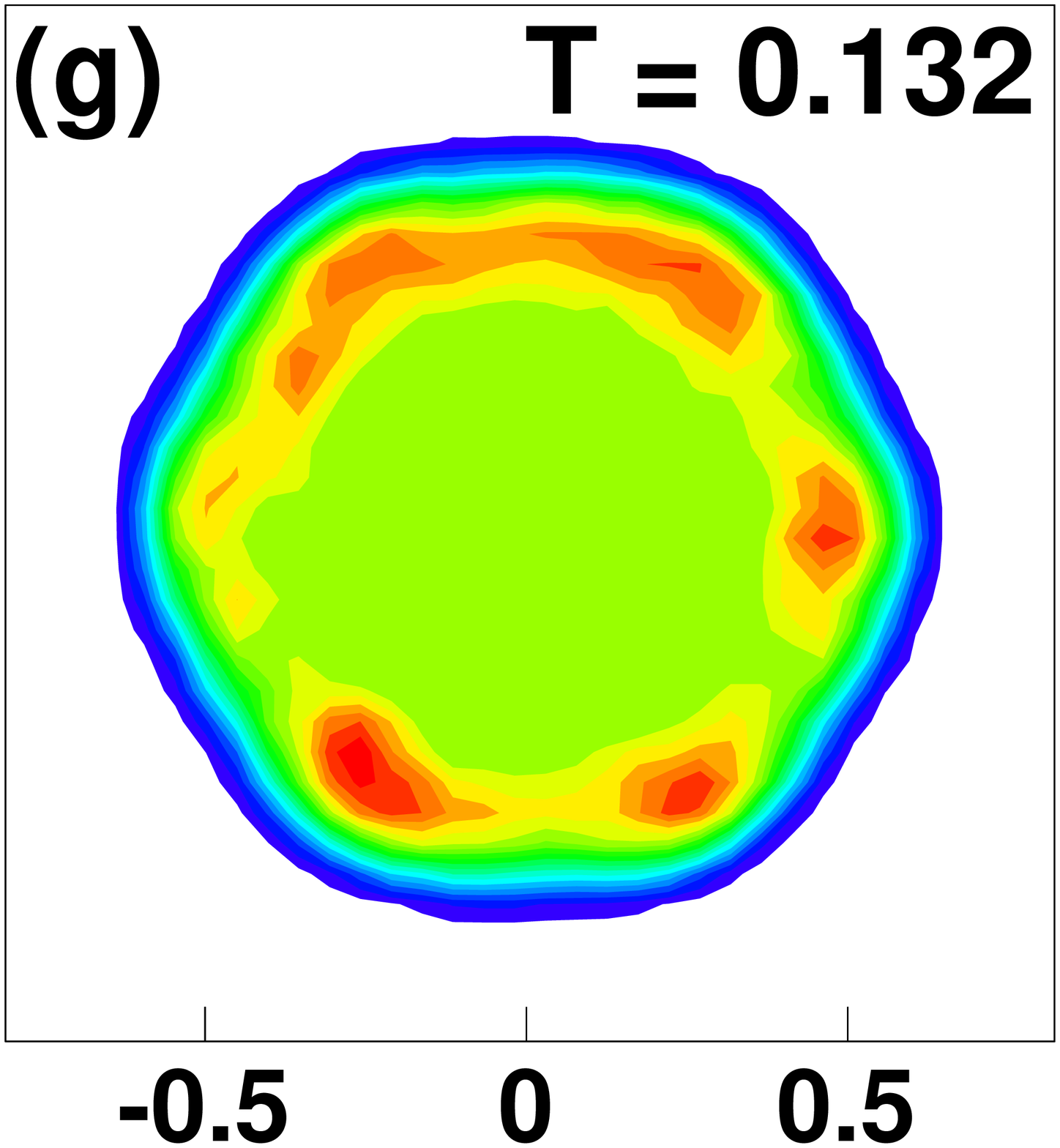}
\includegraphics[width=0.225\columnwidth]{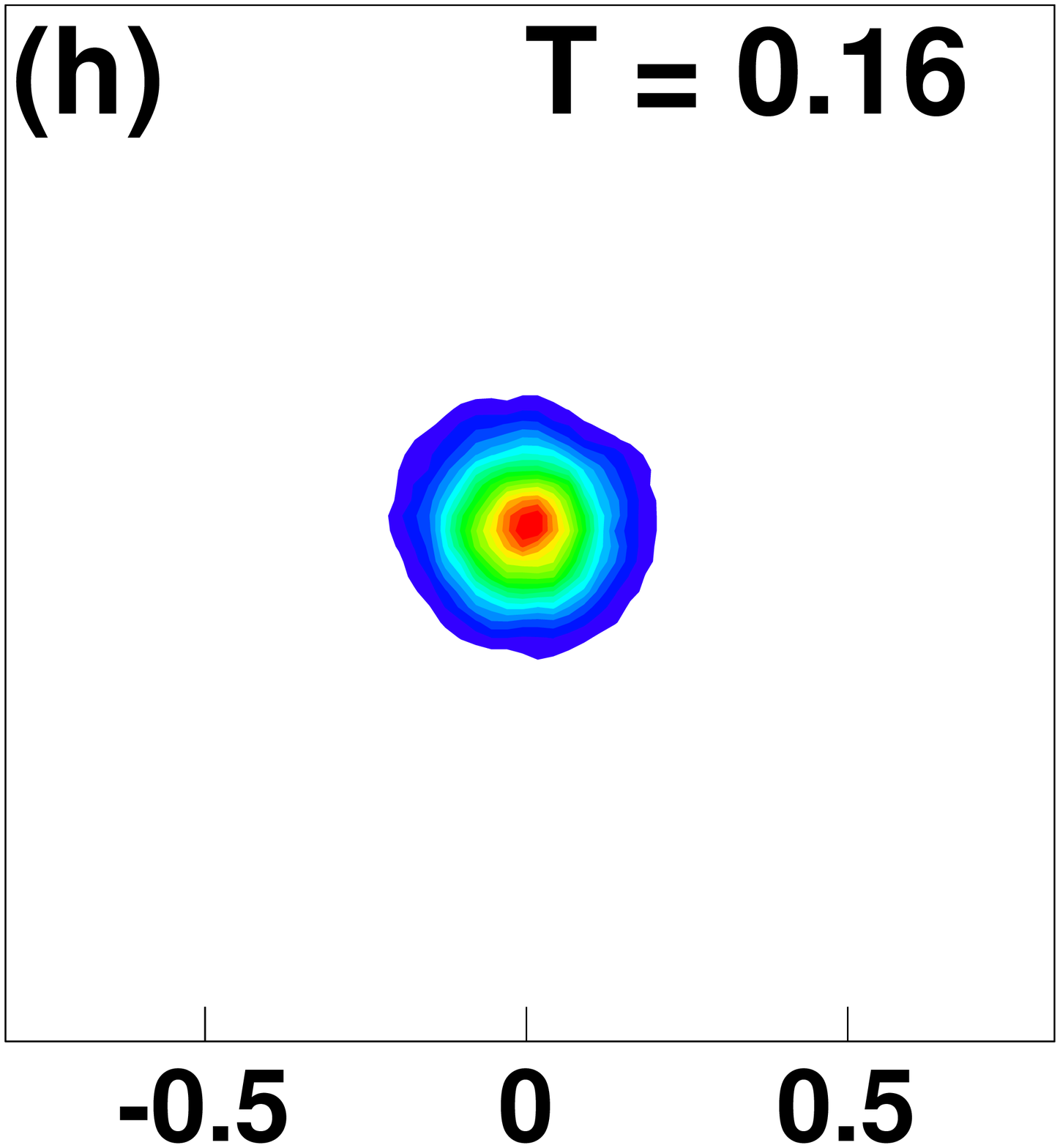}
\caption{
Histograms of the order parameter $m_{N(S)}$, obtained for the system with 2*84*84 spins in the ordered phase, (a) and (e),
in the intermediate phase, (b)-(c) and (f)-(g),
and in the disordered phase, (d) and (h).
Histograms (a)-(d) are  computed for $\alpha=0.25$, and (e)-(h) are  for $\alpha=0.75$.
The histograms are  presented on the complex plane (Re $|m_{N(S)}|$, Im $|m_{N(S)}|$).}
\label{fig:histograms}
\end{figure}
%%%%%%%%%%%%%%%%%%%%%%%%%%%%%%%%%%%%%%%%%%%%%%%%%%%%%%%%%%%%%%%%%%%%%%%%%%%%%%%%

{\it Order by Disorder.}
The symmetry of the KH model combines the cubic symmetry of both the spin  and the lattice space.
It consists  of  simultaneous permutations between the $x, y, z$ spin components and a $C_3$-rotation of the lattice  which defines a discrete symmetry.
The classical ground state has a higher symmetry than that of the Hamiltonian -- the ground state energy does not change under a simultaneous rotation of all  spins.
Since this applies only to the ground state,%and not to  a random spin configuration,
the KH model has only an accidental continuous rotation symmetry.
Its actual symmetry is  discrete; at zero temperature, the "pseudo" SU(2) symmetry is broken by quantum fluctuations that restore the underlying cubic symmetry of the model \cite{jackeli10}.
The magnetically ordered phase is gapped with a spin gap that corresponds to the finite energy cost of deviating the order parameter from one  of the cubic axes.
We  show in the following  that thermal fluctuations of classical spins at finite T also select a collinear spin configuration whose order parameter points along one of the cubic axes.

{\it Parameters of the Simulations.}
We have carried out  classical MC simulations of the model (\ref{ham}) using the standard Metropolis algorithm.
In our MC simulations, we treat the spins as three-dimensional (3D) vectors, ${\bf S}_i=(S_i^x ,S_i^y ,S_i^z)$, of unit magnitude with $(S_i^x)^2 + (S_i^y)^2+ (S_i^z)^2=1$ at every site.
At each temperature, more than $10^7$ MC sweeps  were performed.
Of these, $5*10^5$ were used to equilibrate the system, and afterwards only 1 out of every 5 sweeps was used to calculate the averages of physical quantities.
We  present all energies in the units of $J_H$ and assume  $k_B=1$.
The calculations were carried out on several finite systems with size $2*L*L$  that are spanned by the primitive vectors of a triangular lattice ${\bf a}_1=(1/2,\sqrt{3}/2)$ and ${\bf a}_2=(1,0)$ with a 2-point basis using periodic boundary conditions.
%These clusters  respect all spatial symmetries of the model.

{\it Results.} To study the possible phases of the model (\ref{ham}), we introduce four magnetic configurations (Fig.~\ref{fig:orders}):
a FM order,
a simple two-sublattice AF N\'{e}el order,
a stripy order,
and a zigzag spin order.
The classical energies of these states can be easily computed:
$E_{cl}^{\mathcal{M}}=3-5\alpha$,
$E_{cl}^ {\mathcal{Z}}=-3\alpha+1$,
$E_{cl}^{ \mathcal{S}}=-\alpha-1$,
and  $E_{cl}^{\mathcal{N}}=5\alpha-3$ for the FM, the zigzag, the stripy and the N\'{e}el phases, respectively.
For $0\leq \alpha < 1$, the classical ground state is either the N\'{e}el AF  with the vector order parameter $\mathcal{N}=\frac{1}{N}\sum_{i}({\bf S}_{iA}-{\bf S}_{iB})$ or the stripy phase described  by $\mathcal{S}=\frac{1}{N}\sum_{i=n}({\bf S}_{iA}-{\bf S}_{iB}+{\bf S}_{iC}-{\bf S}_{iD})$.
Here, $A,B$ and $A,B,C,D$   denote either two or four sublattices that respectively characterize  the N\'{e}el AF and stripy order.
The classical phase transition between them occurs  at $\alpha=1/3$.
At $\alpha =1$, the FM, stripy, and zigzag phases all have the same classical energy.
However, the classical degeneracy of this point, which corresponds to  the pure Kitaev model, is much higher.
This limit has  been thoroughly studied by  Baskaran {\it et al.} \cite{baskaran08}.

%%%%%%%%%%%%%%%%%%%%%%%%%%%%%%%%%%%%%%%%%%%%%%%%%%%%%%%%%%%%%%%%%%%%%%%%%%%%%%%%
\begin{figure}
\includegraphics[width=0.9\columnwidth]{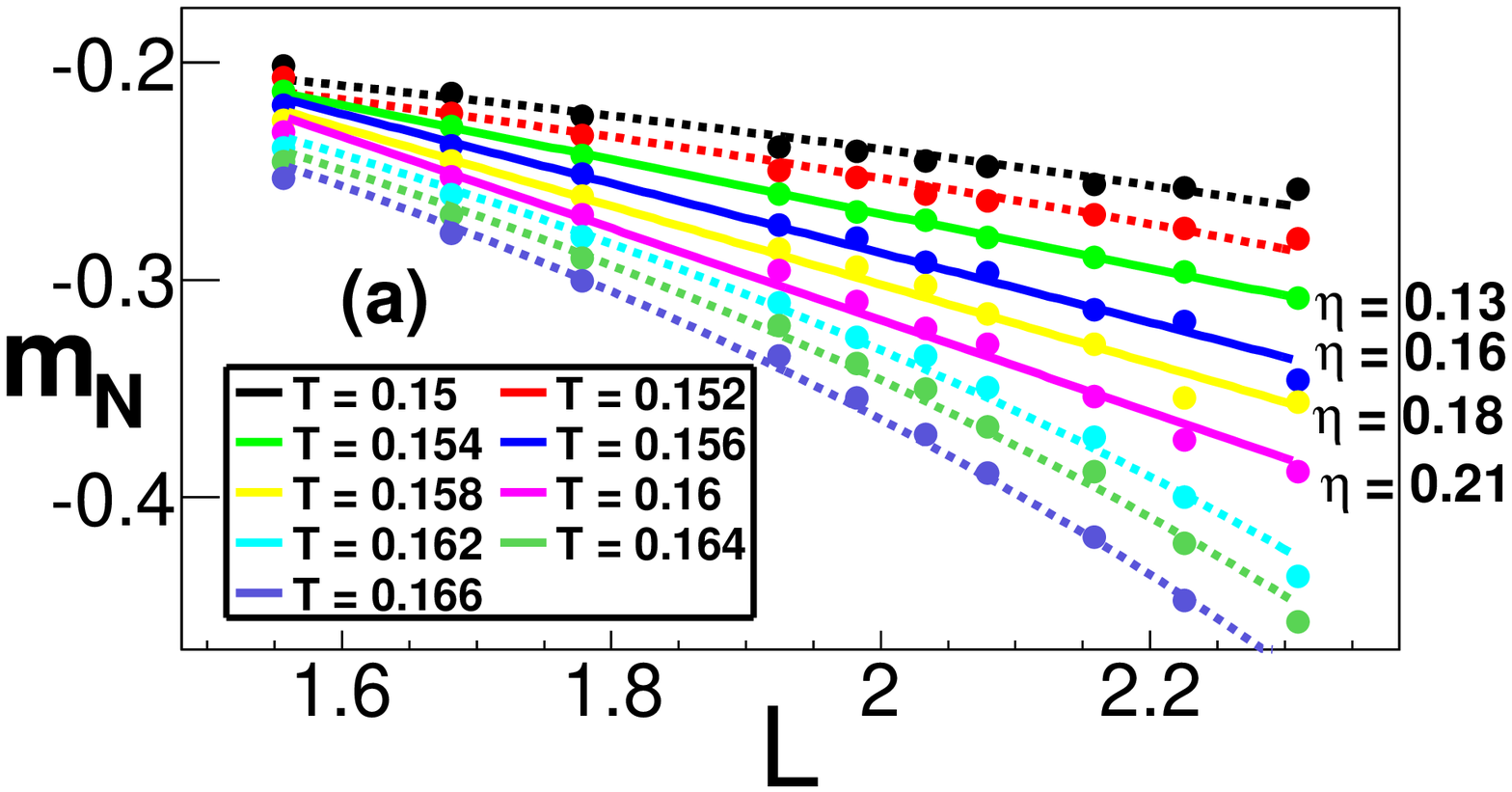}
\includegraphics[width=0.9\columnwidth]{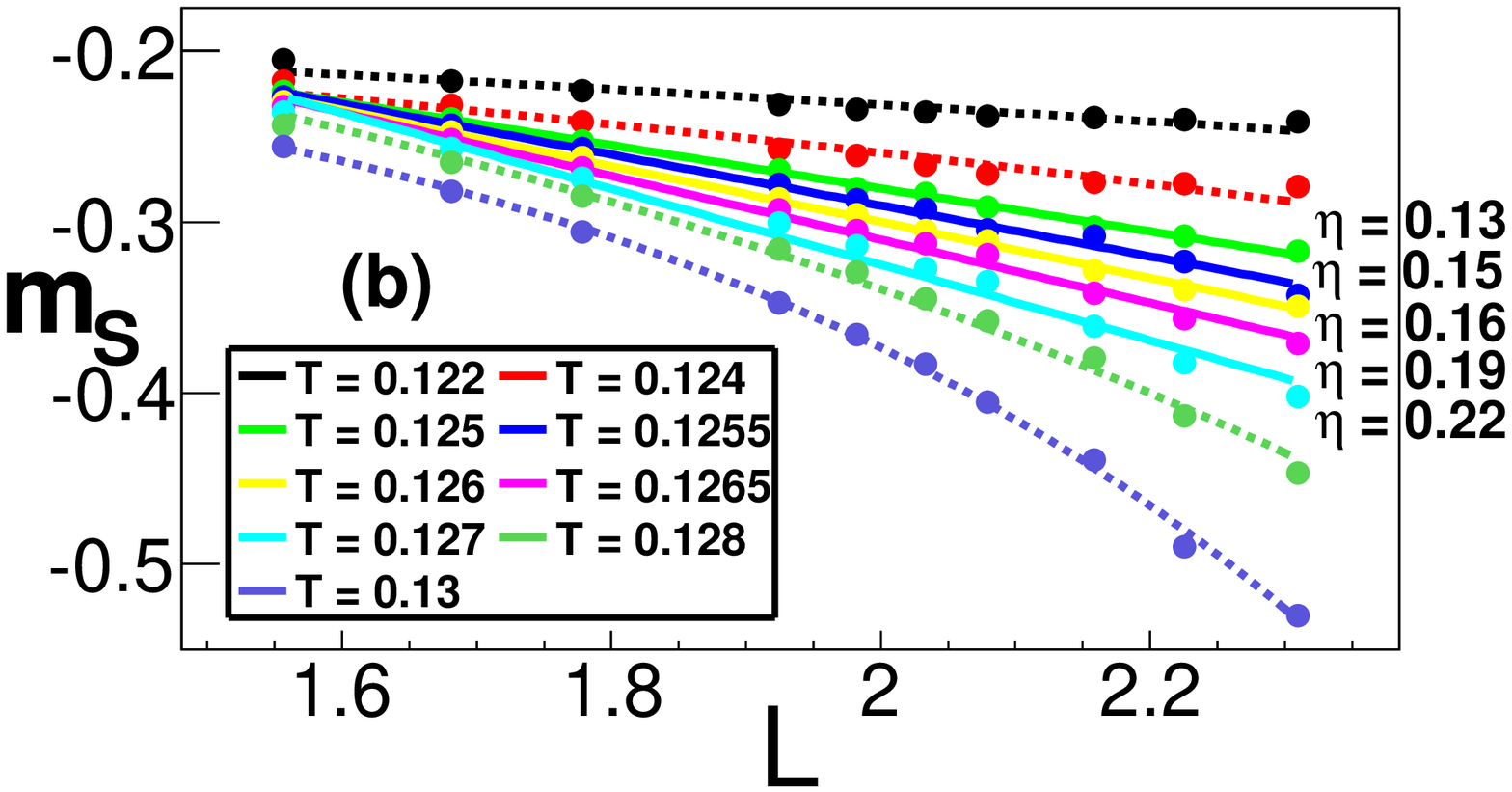}
\caption{
The log-log plots of the order parameter $m_{N(S)}$ as a function of system size $L$ at various temperatures.
The solid curves indicate the linear behavior that corresponds to a power-law dependence, $m_{N(S)}\sim L^{-\eta/2}$,  cooresponding to the intermediate critical phase.
The dashed curves show deviation away from the linear behavior outside the critical phase.
}
\label{fig:loglog}
\end{figure}
\begin{figure}[ht]
\begin{center}
\begin{turn}{90}
\includegraphics[trim=3.12cm 10.18cm 10.20cm 3.12cm,clip=true,keepaspectratio=true,width=5cm]
{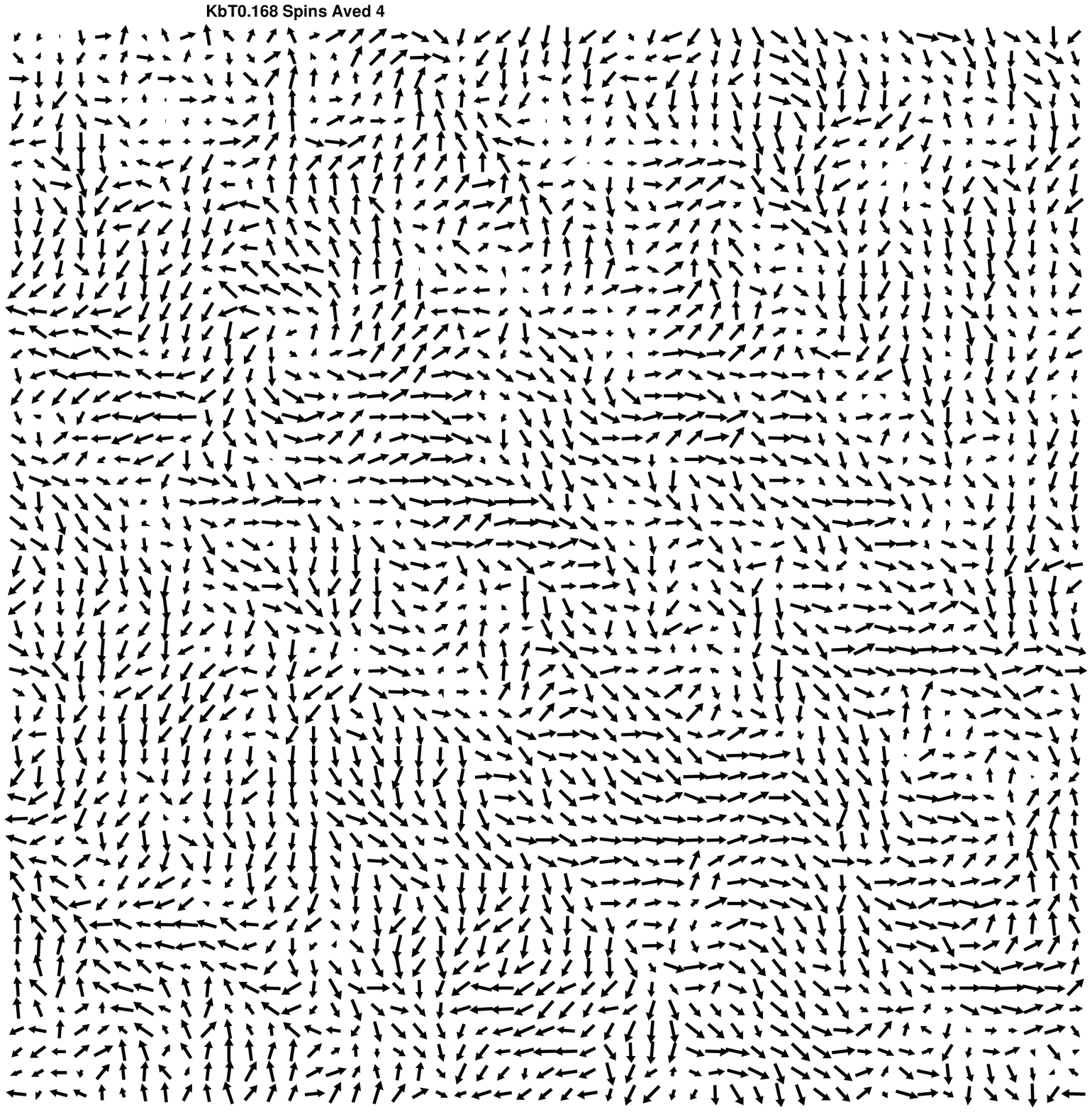}
\end{turn}
\end{center}
\caption[Vortex]{
A snapshot of the coarse-grained order parameter $\langle m_N \rangle$ at $T=0.168$.
The  vortex-like  topological excitations are evident.
}
\label{fig:vor}
\end{figure}
%%%%%%%%%%%%%%%%%%%%%%%%%%%%%%%%%%%%%%%%%%%%%%%%%%%%%%%%%%%%%%%%%%%%%%%%%%%%%%%%

To make an analogy to the six-state clock model, we map the order parameter describing the magnetically ordered phase of the KH model  onto a 2D complex order parameter,
$m_{N(S)}=\sum_{i=1}^6 |m_{i,N(S)}| e^{ \imath\theta_i}$, such that the six possible ordered states  are characterized by  $\theta_i=\pi n_i/3$, $n_i=0,..5$~\cite{chern12}.
The mapping  is  exact only well within the ordered state since there is no guarantee that  the thermal fluctuations of the order parameter will actually have a 2D character given that the  spin degrees of freedom are three-dimensional.
Depending on the strength of the spin stiffness  in different directions, the long-range low-T magnetic order can be destroyed  in one of several ways.
If the stiffness of thermal fluctuations along the circle  is softer than the stiffness of fluctuations in the direction transverse to the circle, the long-range order may be destroyed by a discontinuous first-order transition, by two continuous phase transitions with an intermediate  partially ordered phase, or by two KT phase transitions with an intermediate critical phase \cite{jose77,isakov03,chern12,ortiz12}.
In the last scenario,  the critical phase is destroyed by topological excitations in the form of discrete vortices whose existence is directly related to the emergence of a continuous symmetry; the high-T transition will first bring the system into a disordered phase where fluctuations are primarily 2D, and the crossover to the 3D paramagnet occurs at even higher temperatures.

In Fig.~\ref{fig:histograms} we present the results of the histogram method for the  complex order parameter.
At low temperatures,  Figs.~\ref{fig:histograms} (a) and (e),  a sixfold degeneracy  present in the ordered phase is seen.
For both $\alpha=0.25$ and $\alpha=0.75$, the six states which have the highest weight in the histogram are where the order parameter $m_{N(S)}$ points along  one of the cubic axes.
In Figs.~\ref{fig:histograms} (b) and (f), when the temperature increases  beyond a certain critical temperature, a continuous $U(1)$ symmetry emerges signaling both the disappearance of the sixfold anisotropy and the appearance of the  critical phase.
The formation of vortices can be seen in  Fig.~\ref{fig:vor} where we present a  snapshot of the coarse-grained order parameter $\langle m_N \rangle$ at $T=0.168$.
Upon a further increase in temperature, the amplitude of the order parameter decreases  (Figs.~\ref{fig:histograms} (c) and (g))  until it shrinks to zero indicating the transition to the paramagnetic phase (Figs.~\ref{fig:histograms} (d) and (h)).

To better understand the properties of the intermediate phase and to confirm its critical nature, we performed the finite-size scaling analysis appropriate for KT transitions \cite{challa86}.
The full finite-size scaling analysis is rather involved and will be reported elsewhere \cite{unpublished}.
Here  we present only the scaling behavior of the order parameter.
At the KT transition, the order parameter exhibits the power law dependence on system size, $m\sim L^{-\eta/2}$.
As each point of the intermediate critical phase can be understood as a critical point, the power law behavior of the order parameter should hold throughout the entire  phase.
We  found that the boundaries of the critical phase  are characterized by critical exponents close to $1/9$  and $1/4$ for the lower  and upper boundaries at $T_{c_1}$  and $T_{c_2}$, which is in agreement with critical exponents for the six-state clock model obtained by the renormalization group analysis~\cite{jose77}.
Fig.~\ref{fig:loglog} shows  the log-log plots of the order parameter $m_{N(S)}$  as a function of system size for different temperatures.
For $\alpha=0.25$, the data points  in Fig.~\ref{fig:loglog} a) show a linear behavior in the temperature interval between $T_{c_1}\simeq 0.152$ and $T_{c_2}\simeq 0.162$,
in which  there are several critical lines characterized by $\eta$ between $1/9$ and $1/4$.
For $\alpha=0.75$, we have detected the critical phase in the temperature interval  between $T_{c_1}\simeq 0.125$ and $T_{c_2}\simeq 0.127$.

%Observing the critical behavior proved to be challenging in the case of the KH model.
%Despite the discrete symmetry of the Hamiltonian, the spin degrees of freedom are 3D Heisenberg spins.
%These continuous spins are affected by thermal fluctuations much  more strongly than the  Ising-like spin systems in which  all previous studies of the  six-state clock model's critical phase have been performed.
%  As a result, it requires a larger number of sweeps to sufficiently average out the thermal fluctuations and a sufficiently large system to capture the behavior of the critical phase.
%This translates to a very large amount of CPU time necessary for an accurate study of the critical properties.

%%%%%%%%%%%%%%%%%%%%%%%%%%%%%%%%%%%%%%%%%%%%%%%%%%%%%%%%%%%%%%%%%%%%%%%%%%%%%%%%
\begin{figure}
\includegraphics[width=0.45\columnwidth]{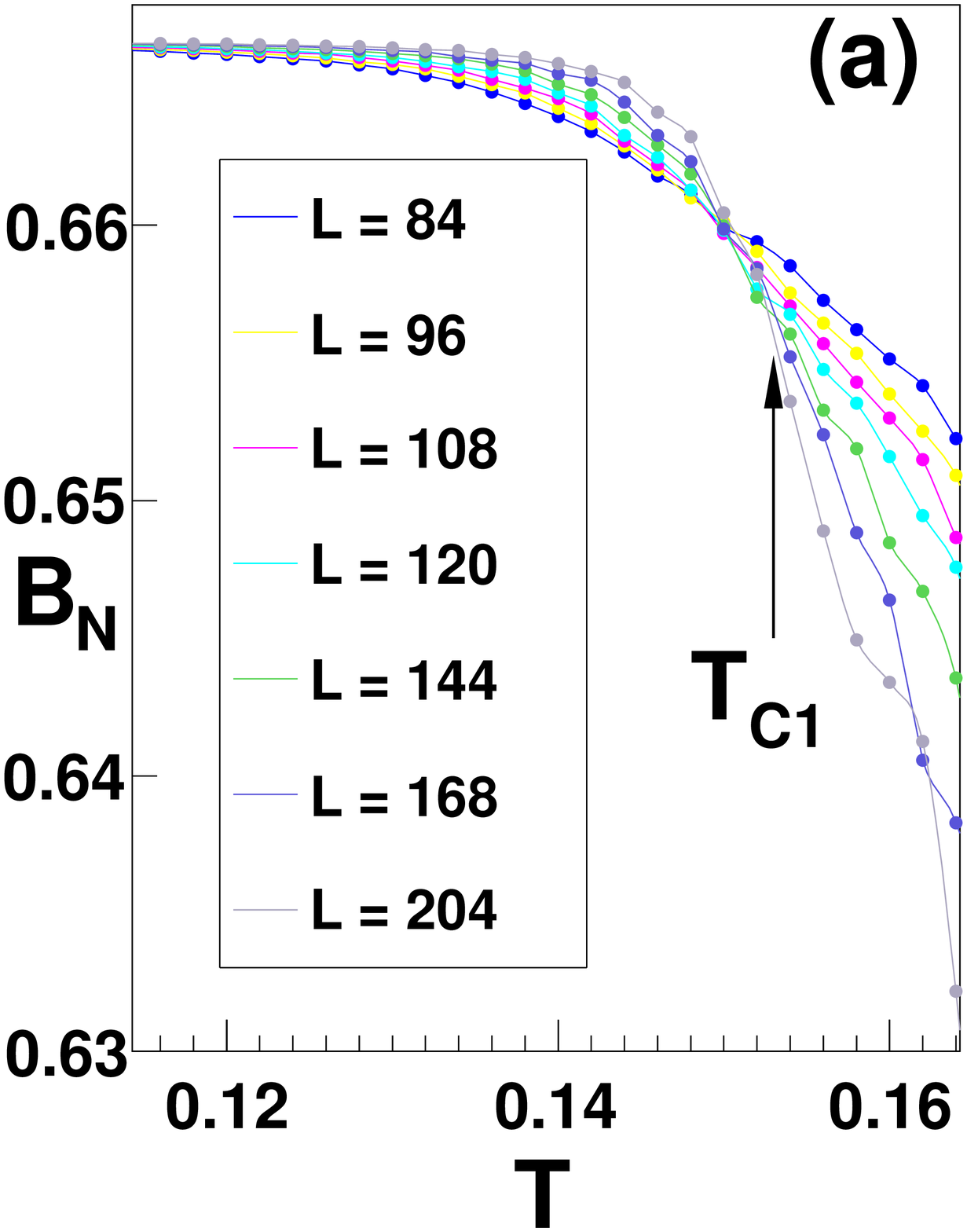}
\includegraphics[width=0.45\columnwidth]{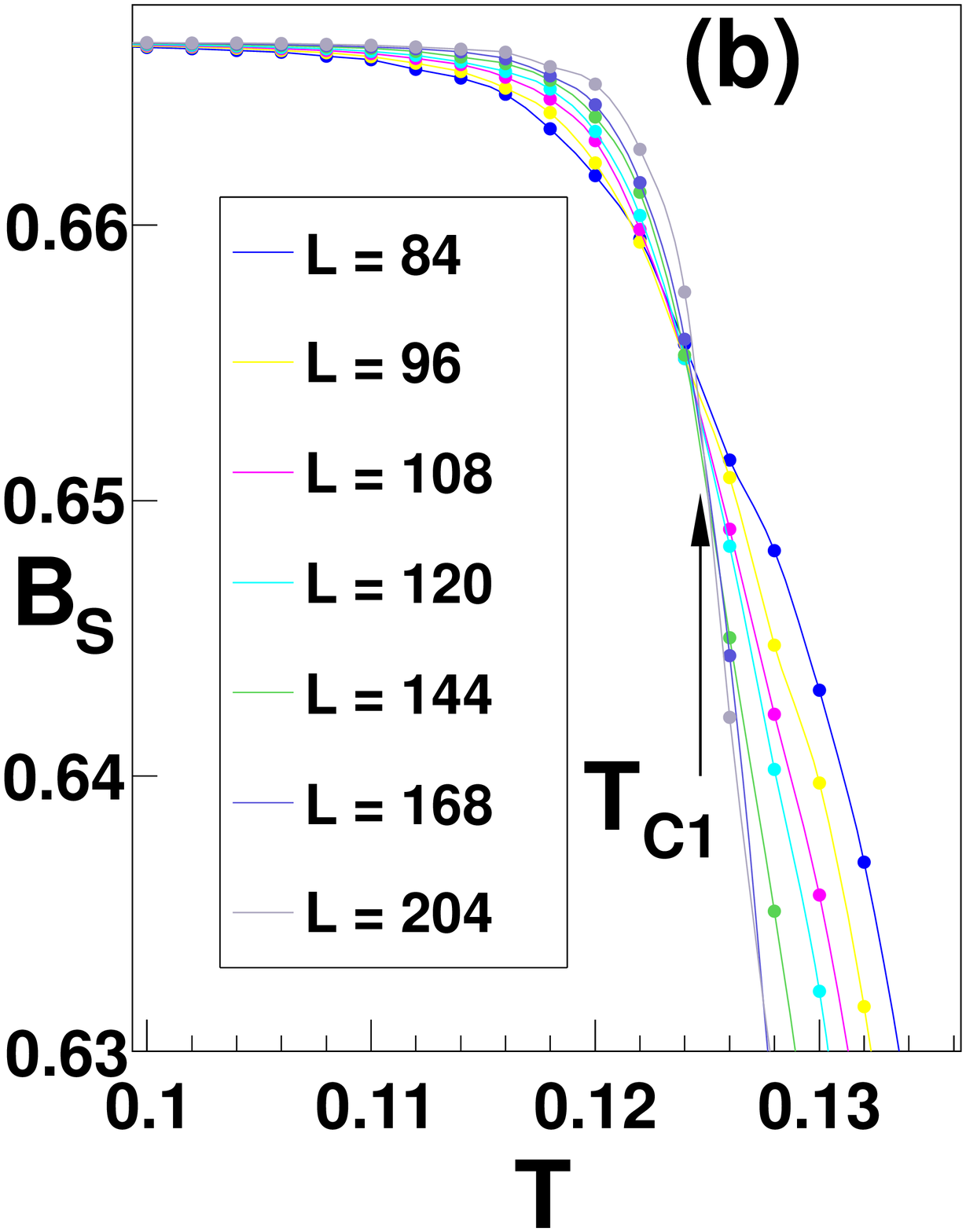}
\caption{
The Binder cumulant as a function of temperature for (a) $\alpha=0.25$ and (b) $\alpha=0.75$.
From  the  crossing points of different Binder's curves, we estimate $T_{c_1}=0.152$ and  $T_{c_1}=0.124$ for  $\alpha=0.25$ and  $\alpha=0.75$, respectively.
}
\label{fig:binder}
\end{figure}
%%%%%%%%%%%%%%%%%%%%%%%%%%%%%%%%%%%%%%%%%%%%%%%%%%%%%%%%%%%%%%%%%%%%%%%%%%%%%%%%

The lower transition temperature $T_{c_1}$ can  be independently determined using fourth-order Binder cumulant (Figs.~\ref{fig:binder} (a) and (b)).
The Binder cumulant has a scaling dimension of zero; thus the crossing  point of the cumulants for different lattice  sizes  provides a reliable estimate for the value of the critical temperature $T_{c_1}$ at which the long range order is destroyed.
The crossing points for $\alpha=0.25$ and $\alpha=0.75$ are $T_{c_1}=0.152$ and $T_{c_1}=0.124$, respectively.
They are in good agreement with estimates obtained from the log-log plots in  Fig.~\ref{fig:loglog}.

\begin{figure}
\includegraphics[width=0.95\columnwidth]{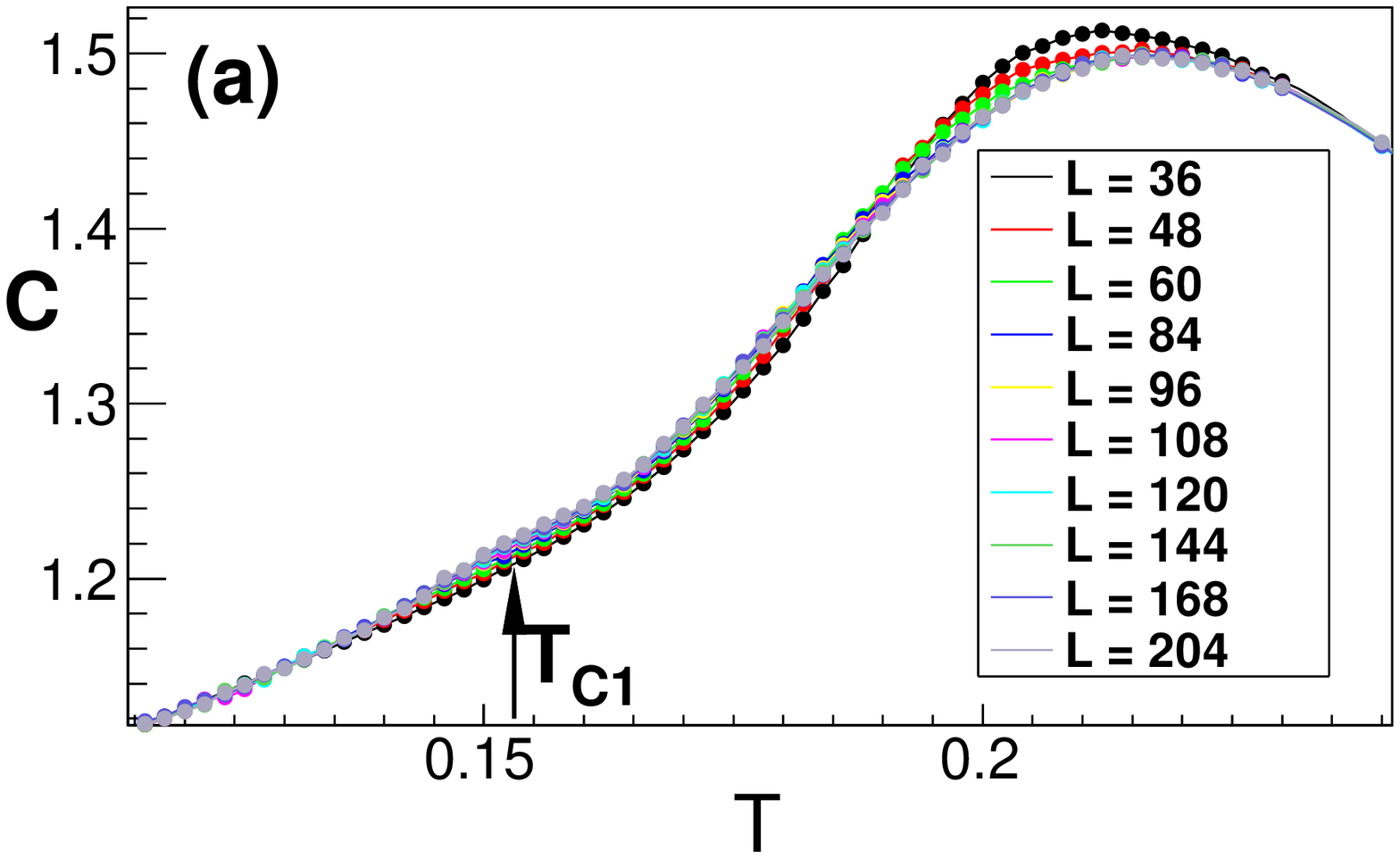}
\includegraphics[width=0.95\columnwidth]{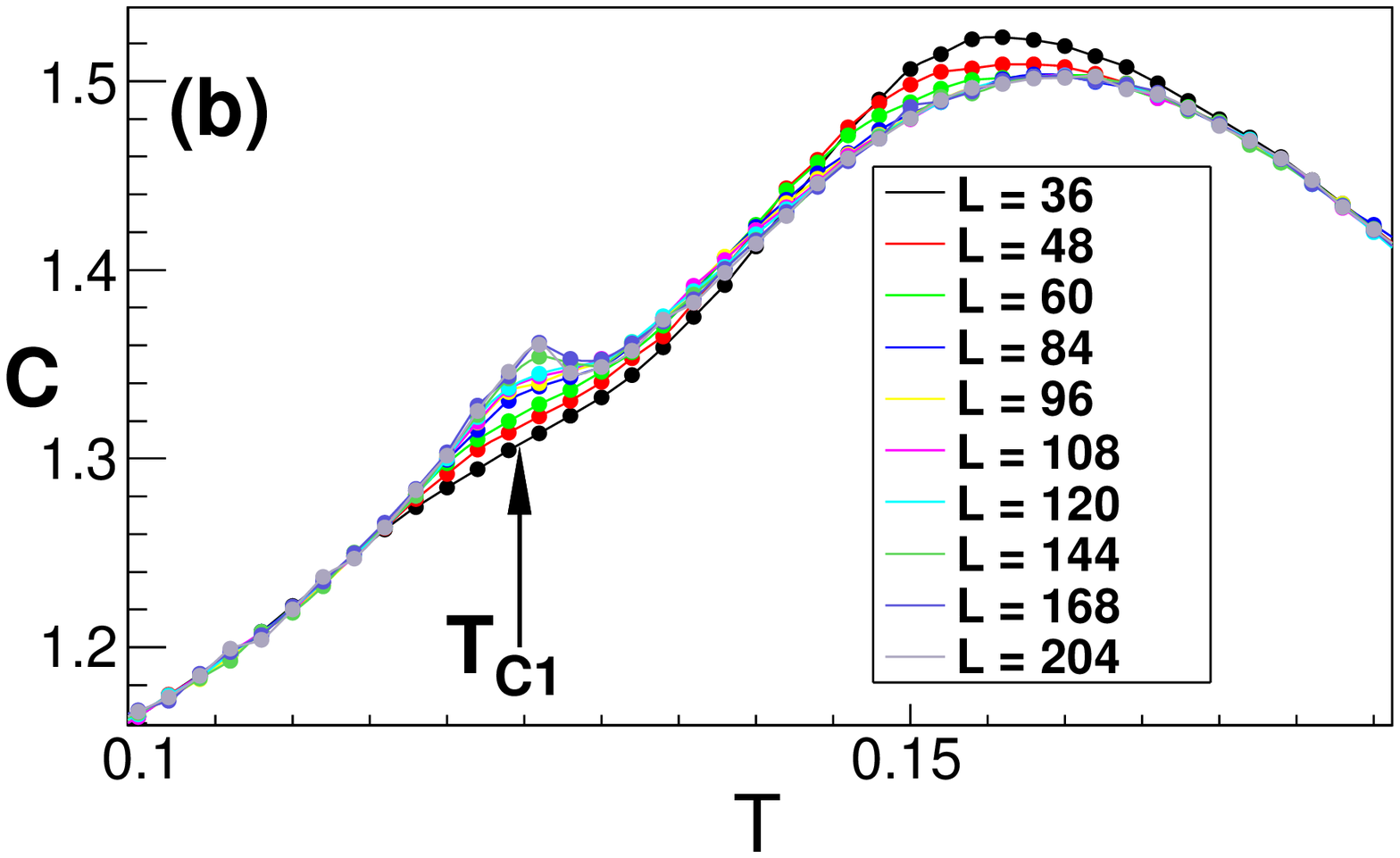}
\caption{
Specific heat $C$ as a function of temperature for (a) $\alpha=0.25$ and (b) $\alpha=0.75$.
}
\label{fig:specific}
\end{figure}

In Figs.~\ref{fig:specific} (a) and (b) we present the temperature dependence of the specific heat, $C=(\langle E^2\rangle-\langle E\rangle^2)/NT^2$. While the low-T transition, seen as small peak at temperatures $T_{c_1} =  0.152$  and $0.1247$ for $\alpha=0.25$ and $0.75$, respectively, is in a good agreement with our previous estimates,  the features corresponding to the high-T transition $T_{c_2}$ are barely distinguished by eye. This is not surprising as
the high-T transition is a usual KT
transition at which the specific heat does not diverge at
the critical point~\cite{last}. It is also likely that the high-T KT
transition might be shadowed by the crossover to the 3D
paramagnet, which is seen in Fig.~\ref{fig:specific}
as a very broad hump at higher-T.

%%%%%%%%%%%%%%%%%%%%%%%%%%%%%%%%%%%%%%%%%%%%%%%%%%%%%%%%%%%%%%%%%%%%%%%%%%%%%%%%
%%%%%%%%%%%%%%%%%%%%%%%%%%%%%%%%%%%%%%%%%%%%%%%%%%%%%%%%%%%%%%%%%%%%%%%%%%%%%%%%

Our findings for the specific heat show a lot of similarities between the  experimental data obtained on the Na$_2$IrO$_3$  and  Li$_2$IrO$_3$ compounds by
Refs.~\cite{gegenwart10,gegenwart12} and \cite{takagi11}.
%Save space by not mentioning names?
%Singh and  Gegenwart \cite{gegenwart10,gegenwart12}, and by the group of Takagi \cite{takagi11}.
%We need to make sure we are citing the correct paper here.
In  Na$_2$IrO$_3$,  both the lambda-like anomaly  at the N\'{e}el ordering temperature, $T_N=15$ K,  and a broad tail which extends into higher  temperatures are seen in the specific heat measurements~\cite{gegenwart10}.
The latter suggests the presence of short-range order above the bulk 3D ordering that can be understood by our proposed scenario of the critical phase.

Let  us  estimate the temperatures of the KT transitions and  the  width of the critical phase in Na$_2$IrO$_3$ based on our results obtained for  the KH model with $\alpha=0.25$.
On the mean field level, the exchange on the NN bonds may  be estimated from the classical energy,  $J_1\simeq (3-5\alpha)/3$,  in the N\'{e}el phase.
 From the recent neutron scattering experiment
% by Choi {\it et al.}
\cite{choi12}, the NN exchange in Na$_2$IrO$_3$ was estimated to be $J_1=4.17$ meV.
In the bulk of our paper,  all energies were measured in the units of $J_H$, and thus we estimate $J_1$ to be equal to 12.7 meV.
 This gives the prediction for the critical temperature  to be $T_{c_1}=16.8$ K, which is very close to  the experimental value $T_N=15$ K \cite{gegenwart10,gegenwart12}.
%0.152*4.17*4*11.6/1.75=16.8 K
%(3-5alpha)/3*S^2=(3-1.25)/3=1.75/4 should be 4.17. Thus our energy unit is 12.7 meV
 Our estimate for the upper boundary of the critical phase is $T_{c_2}=17.7$ K which makes the predicted critical phase very narrow.
 %Due to the lack of neutron scattering data for Li$_2$IrO$_3$, it is difficult to estimate the width of the critical phase for this compound, but   we expect it to be  narrow too.
  We note here that the critical phase  survives  in the extended KH model with included further-neighbor exchange couplings~\cite{choi12,gegenwart12,mazin12}  which  are essential for  comparison with experiment.
 However,  in order to determine the upper boundary of  the critical phase   additional extensive numerical simulations must be performed.

{\it Acknowledgements.}
The authors are particularly thankful  to C. Batista, G.-W. Chern, G. Jackeli, and Y. Kato for stimulating discussions and many helpful suggestions.
We are grateful to H. Takagi and T. Takayama for sharing with us unpublished data on Na$_2$IrO$_3$ and Li$_2$IrO$_3$.
N.P. acknowledges the support from NSF grant DMR-1005932.
N.P. also thanks the hospitality of the visitors program at MPIPKS, where the part of the work  has been done.

%%%%%%%%%%%%%%%%%%%%%%%%%%%%%%%%%%%%%%%%%%%%%%%%%%%%%%%%%%%%%%%%%%%%%%%%%%%%%%%%

%%%%%%%%%%%%%%%%%%%%%%%%%%%%%%%%%%%%%%%%%%%%%%%%%%%%%%%%%%%%%%%%%%%%%%%%%%%%%%%%

\end{document}